**Towards a Perceptual Distance Metric for Auditory Stimuli**

Sarah Oh[1], Elijah FW Bowen[1], Antonio Rodriguez[1], Damian Sowinski[2], Eva Childers[1], Annemarie Brown[1], Laura Ray[2], Richard Granger[1,2]

[1] Brain Engineering Lab, Department of Psychological & Brain Sciences, Dartmouth College, Hanover, New Hampshire 03755, USA

[2] Thayer School of Engineering, Dartmouth College, Hanover, New Hampshire 03755, USA

**Abstract**
Although perceptual (dis)similarity between sensory stimuli seems akin to distance, measuring the Euclidean distance between vector representations of auditory stimuli is a poor estimator of subjective dissimilarity. In hearing, nonlinear response patterns, interactions between stimulus components, temporal effects, and top-down modulation transform the information contained in incoming frequency-domain stimuli in a way that seems to preserve some notion of distance, but not that of familiar Euclidean space. This work proposes that transformations applied to auditory stimuli during hearing can be modeled as a function mapping stimulus points to their representations in a perceptual space, inducing a Riemannian distance metric. A dataset was collected in a subjective listening experiment, the results of which were used to explore approaches (biologically inspired, data-driven, and combinations thereof) to approximating the perceptual map. Each of the proposed measures achieved comparable or stronger correlations with subjective ratings ($r \approx 0.8$) compared to state-of-the-art audio quality measures.

**I.  Introduction**

In perception, physical stimuli reach sensory receptors and are transduced into electrical impulses which are in turn processed by increasingly specialized and complex brain circuits. Physical stimuli enter the brain as an array of features whose dimensions are orthogonal, meaning it is possible to vary any one component without affecting the others. Subsequent processing stages build upon combinations of these features. The eventual result is a subjective perceptual representation. The principle driving this work is the idea that although perceptual dissimilarity seems akin to distance, the familiar Euclidean metric is a poor estimator of dissimilarity. We stipulate that stimulus and perceptual space are Riemannian manifolds, and that the mapping from the stimulus to the perceptual manifold induces a Riemannian metric tensor that can better measure perceptual distances between stimulus points.

Suppose $m$-dimensional manifold $\mathcal{M}$ is the stimulus manifold, and the $n$-dimensional manifold $\mathcal{N}$ is the perceptual manifold. $\mathcal{M}$ is assigned the standard orthonormal coordinate basis in $\mathbb{R}^m$, defining $\mathcal{M}$ as a subset of Euclidean space $\mathbb{E}^m$ and the metric $g_\mathcal{M}$ defining the inner products on the tangent spaces of $\mathcal{M}$ as the dot product. The line element

$$ds_{g_\mathcal{M}}^2 = \sum_{i,j} dx_i \, g_\mathcal{M}\left(\frac{\partial}{\partial x_i}, \frac{\partial}{\partial x_j}\right) dx_j = \sum_{i,j} dx_i \, g_{\mathcal{M}_{ij}} dx_j$$

where $\frac{\partial}{\partial x_i}$ are coordinate basis vectors of $\mathbb{R}^m$, becomes

$$ds_{g_{\mathcal{M}}}{}^2 = \sum_{i,j} dx_i\, \delta_{ij} dx_j = \overrightarrow{dx}^T \overrightarrow{dx}$$

Assigning standard coordinates to points in $\mathcal{N}$ similarly gives

$$ds_{g_{\mathcal{N}}}{}^2 = \overrightarrow{dy}^T \overrightarrow{dy}$$

Given a smooth map $f: \mathcal{M} \to \mathcal{N}$ the Jacobian matrix at a point $p \in \mathcal{M}$ is computed by taking the partial derivatives of the coordinates $y(f(p)) = (y_1, \dots, y_n) \in \mathbb{R}^n$ of $f(p)$ with respect to the coordinates $x(p) = (x_1, \dots, x_m) \in \mathbb{R}^m$ of $p$. This matrix relates differentials $\overrightarrow{dx}$ at $p$ to differentials $\overrightarrow{dy}$ at $f(p)$. The metric tensor in $\mathcal{N}$ can be "pulled back" into $\mathcal{M}$ by expressing differentials in $\mathcal{N}$ with respect to differentials in $\mathcal{M}$ via the Jacobian matrix $J$:

$$dy_i{}^2 = \sum_j \frac{\partial y_i}{\partial x_j} dx_j$$

$$\therefore ds_{g_{\mathcal{N}}}{}^2 = \left(J\overrightarrow{dx}\right)^T \left(J\overrightarrow{dx}\right) = \overrightarrow{dx}^T J^T J \overrightarrow{dx} = \overrightarrow{dx}^T g'_{\mathcal{M}} \overrightarrow{dx} = ds_{g'_{\mathcal{M}}}{}^2$$

The term $J^T J = g'_{\mathcal{M}}$ is the pullback of the metric tensor matrix in $\mathcal{N}$ to $\mathcal{M}$ and defines a new (non-Euclidean) metric tensor in $\mathcal{M}$ – which we call the perceptual metric tensor. The line integral using this new metric

$$s_{g'_{\mathcal{M}}} = \int_{t_1}^{t_2} \sqrt{ds_{g'_{\mathcal{M}}}{}^2} = \int_{t_1}^{t_2} \sqrt{\sum_{ij} \frac{dx_i}{dt} g'_{\mathcal{M}\,ij} \frac{dx_j}{dt}}\, dt$$

measures the length $s_{g'_{\mathcal{M}}}$ of the curve in $\mathcal{M}$ with coordinates $\gamma(t) = (x_1(t), \dots, x_m(t))$ in $\mathbb{R}^m$, as the length of the mapped curve in $\mathcal{N}$. The Riemannian distance $d_{g'_{\mathcal{M}}}(r, s)$ computed by this metric between points $r, s \in \mathcal{M}$ is the infimum length of all curves in $\mathcal{M}$ connecting $r$ and $s$. Because manifolds are locally Euclidean, given $\mathcal{M}, \mathcal{N}, f$, and points $p, q \in \mathcal{M}$ that are close enough together, the points in the neighborhood of $p$ and $q$ will have approximately the same Jacobian; i.e., the relationships between displacements of coordinates $y_i$ in $\mathbb{R}^n$ and displacements of coordinates $x_i$ in $\mathbb{R}^m$ remain roughly constant along the infimum length curve. The curve from $f(p)$ to $f(q)$ thus approximates a straight line, allowing us to use the approximation

$$s_{g'_{\mathcal{M}}} \approx \|y(f(p)) - y(f(q))\| \tag{1}$$

Our geometric framework is illustrated in Figure 1.

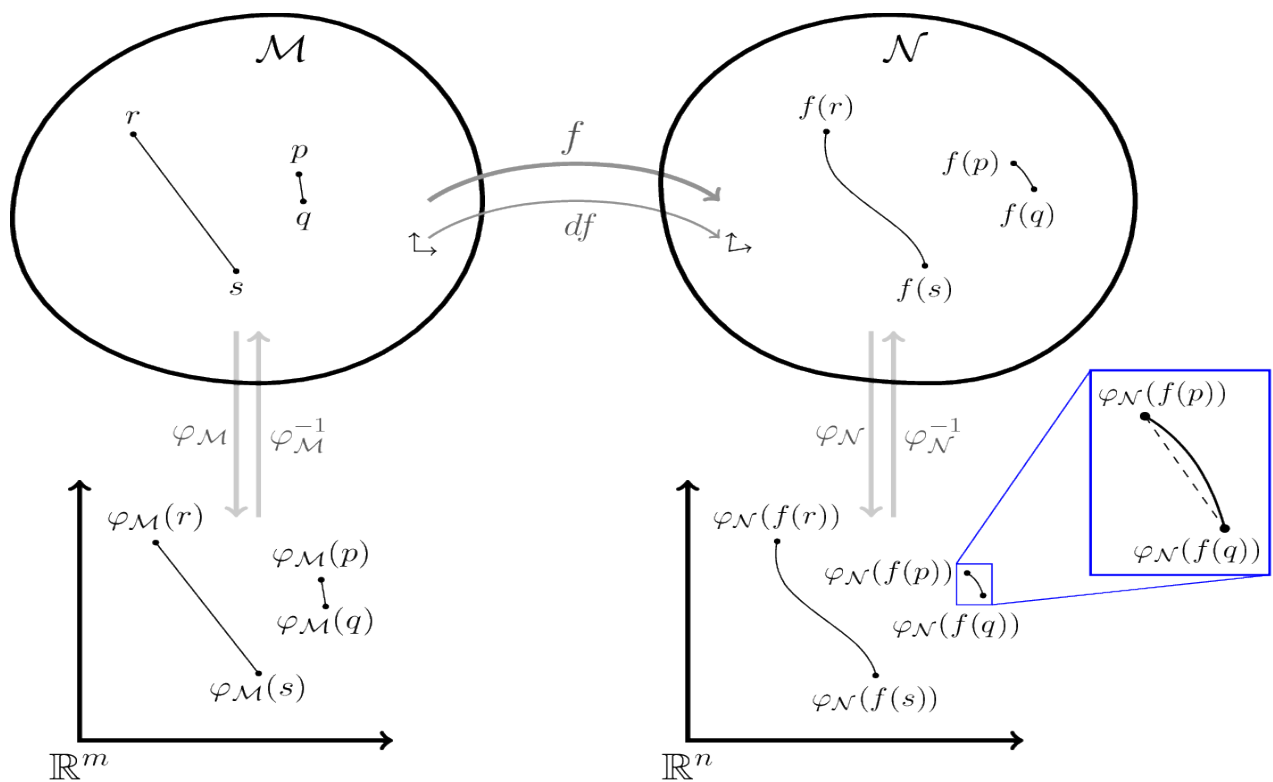

*Figure 1* Illustration of geometric framework for perceptual distance. $f$ maps points on stimulus manifold $\mathcal{M}$ to points on perceptual manifold $\mathcal{N}$. $x$ and $y$ assign local coordinates in $\mathbb{R}^m$ and $\mathbb{R}^n$ to points in $\mathcal{M}$ and $\mathcal{N}$, respectively (axes drawn to evoke a sense of Cartesian coordinates, not to indicate number of dimensions). $df$ maps vectors in the tangent space of $\mathcal{M}$ to vectors in the tangent space of $\mathcal{N}$ via the Jacobian matrix at each point. The zoomed-up region illustrates how the curve mapped from a very short curve in $\mathcal{M}$ approximates a straight line.

The goal of this work is to objectively measure perceived dissimilarity between auditory stimuli by estimating the Riemannian distances between stimulus points. The task of objective dissimilarity measurement is only useful, and possible, in cases where subjects tend to agree with each other about dissimilarity. Top-down processing – perception influenced by cognition – causes broad variability across individuals and contexts [1] [2]. Bottom-up processing, by contrast, is stimulus-driven and thus less dependent on observer and context. When auditory stimuli are drawn from different sources, top-down processing inevitably plays a larger role in comparisons of auditory stimuli from different sources because subjects are free to employ cognition to weigh import on various features (e.g. timbre, tonality, volume). To control for some of these sources of variability, we focus on the task of predicting subjective evaluations of dissimilarity between short (1.5) audio clips degraded (to various levels) one audio format, MPEG-1 Layer 1 [3].

For this work, we process PCM audio clips into a series of stimulus vectors for overlapping windows, where vector components are the powers of frequency bands corresponding to evenly spaced locations along the basilar membrane (BM) within the cochlea. See Methods for details. The perceptual distance between two clips is computed as the average distance measured using our approximation between corresponding stimulus vectors. We assume that a degraded clip and its undegraded source clip generally have stimulus vectors that are similar enough to justify our use of the distance approximation given in Equation 1.

## II. Methods
## 1. Subjective Listening Experiment

   a. Clip Selection
   Five classical composers were selected for this study: Chopin, Beethoven, Mozart, Haydn and Bach. 100 tracks from each composer were directly ripped from CDs to obtain a 16-bit PCM-encoded audio waveform with a 44.1 kHz sampling rate. 1.5-second clips from each track were selected at 10 second intervals. Stereo was converted to mono using the `ffmpeg` script [9] and RMS loudness was normalized within each selection to -23dB using the `ffmpeg_normalize` script. 3 segments were selected from each track, ensuring that there was a variety of musical phrasing and that the segments did not start or end mid-note.
   The 480 selected clips were each degraded using a custom implementation of MPEG Layer I [3] at bitrates 40, 32, 24, and 16 kbps, resulting in a total of 2400 clips including the reference (undegraded) clips.

   b. Subjects
   36 normal-hearing subjects participated in this study. Ages varied from 18 to 53, with the average age being 25 years. All subjects were screened to have no hearing impairments or tone deafness. Each subject signed a written consent form that was approved by the Committee of Protection of Human Subjects at Dartmouth College. It would be infeasible for each subject to listen to all 2400 comparisons, so the subjects were split into groups of 12. Within a group, each subject was asked to compare 200 randomly selected clips to the undegraded reference clip, resulting in a single rating given to each of the 2400 comparisons. 3 groups were tested for this experiment, resulting in a total of 3 ratings given to each comparison.

   c. Audio Equipment
   All subjects used headphones that had a frequency response of 15 Hz - 20 kHz. Each subject was able to adjust the volume settings on the headphones to their preference before starting the experiment.

   d. Experimental Procedure
   Subjects listened to both the original and (possibly) degraded version of the short 1.5 second music segment during each round. Each subject used a rating bar located below the continuous rating bar to provide their quality judgment. The ends of this continuous rating bar were labeled with "0" and "100" – "0" meaning the test clip sounded completely degraded from the reference clip, "100" meaning the test clip sounded identical to the reference clip. Once the subject selected a score to rate the degraded version and pressed the "Click Line" button, they were presented with the next set of music segments. See Supplementary Methods for the instructions and rating interface provided to subjects. The experiment is based on the ITU-R recommendation for the subjective assessment of intermediate quality levels of coding systems, or MUlti Stimulus test with Hidden Reference and Anchor (MUSHRA) [10], disregarding the recommendation that only expert listeners be used as subjects.

Each subject was presented with 3 practice rounds that were selected by the authors as being a practical representation of the music segment qualities that would be encountered during the study. Every subject completed 200 rounds after the 3 practice rounds. They were given a short break after every 40 rounds. This study was followed by a quick survey session, which involved the subject answering short questions using rating bars.

The final dataset consists of 480 reference audio clips and the 4 degradations of each, with 3 ratings given by randomly assigned subjects to each degradation pair. The mean ratings given by users to each degradation level are shown with statistics in Supplementary Discussion.

## 2. Data Preprocessing

In early auditory processing, sound waves exert pressure on the eardrum and vibration is transferred to inner ear fluid, producing travelling waves along the basilar membrane (BM) within the cochlea. The properties of the BM (width, stiffness, etc.) vary along its length and determine the characteristic frequency (the frequency that causes the greatest vibration at a particular location along the membrane) at each point. The hair cells of the Organ of Corti convert BM vibration into nerve impulses, which are sent via the auditory nerve to be processed by the auditory cortex [11]. To measure the distances between clips using the proposed metric, they are preprocessed into a representation resembling the physical stimulus received by the auditory system. For each clip, the sound pressure of different frequencies reaching the BM throughout are approximated as follows and as illustrated in Figure 2.

Each signal is first divided by its maximum value to normalize the power spectrum to a 0 dB maximum. Overlapping windows of 1024 time-domain samples are taken with 50% overlap to analyze short frames of the signal as stimulus points. These are multiplied by the tapered Hann window to reduce the effects of spectral leakage introduced by truncating the signal into short segments, and the power spectrum of each windowed frame is computed. Given the 44.1 kHz sampling rate of the audio used, this corresponds to a frequency resolution of 43 Hz and a temporal resolution of 23.2 ms; this is chosen based on findings that the frequency resolution of human hearing can reach about 20 Hz [12], and temporal resolution can reach less than 10 ms [13].

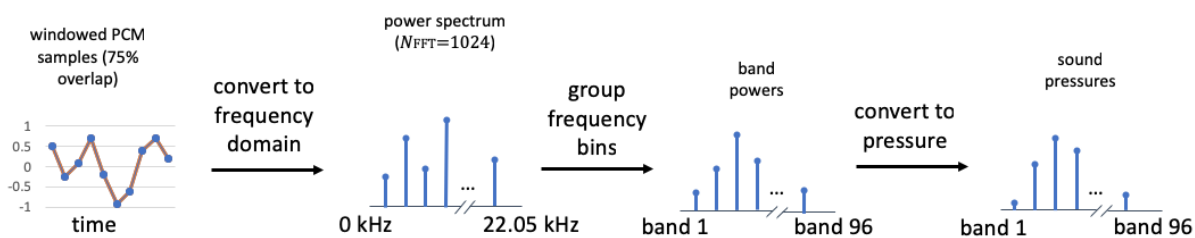

*Figure 2* Preprocessing steps for a single 1024-sample frame of an input clip

Rather than grouping the spectral lines into frequency bands using perceptual frequency scales based on psychoacoustic experiments (e.g., Bark scale, ERB scale), we use the frequency-position relationship for the human cochlea:

$$F = A(10^{ax} - k)$$

where $F$ is frequency, $x$ is a characteristic frequency (CF) position on the BM, $A = 165.4$, $a = 2.1$ (if $x$ is expressed in proportion of total BM length), and $k = 0.88$ [14], to group power spectral lines into 96 frequency bands corresponding to evenly spaced locations on the cochlea and compute the average power in each band.

Sound pressure, measured in Pascals, is the deviation between local and ambient pressure caused by a sound wave. Sound Pressure Level (SPL) is a logarithmic measure for the sound pressure relative to a 20 μPa (serving as the threshold of human hearing) reference level. The SPL of each frequency band is computed using the formula used by the MPEG-1 standard to estimate SPL from a normalized power spectrum $P$:

$$SPL = PN + 10 \log_{10} P$$

[15]. The power normalization term $PN$, which becomes the maximum SPL present in the signal, is set to 90.302 dB – somewhere in between normal conversation (around 70 dB SPL) and loud rock music (100 dB SPL) [16] – assuming playback levels are unknown. SPL is then converted to sound pressure (Pa) using $SP = 20^{-6} 10^{SPL/20}$.

Each clip is thus preprocessed into 64 vectors (one for each windowed frame) representing the effective sound pressures in different frequency bands, each component stimulating a different region along the BM. This ignores interactions among stimulus components that act over time and assumes that input components only affect output components occurring in the same windowed frame. Despite this room for future work, using the proposed metrics to measure distances between corresponding sound pressure frames from the reference and degraded clips gives predictions of the subjective ratings that are comparable with or outperform predictions given by state-of-the-art methods.

## 3. Proposed Perceptual Distance Measures

For each of the proposed modeling methods, the distance between two stimulus clips is computed as the average perceptual distance measured between stimulus point pairs from corresponding frames:

$$\text{distance}(\text{ref}, \text{test}) \triangleq \frac{1}{N_{\text{frames}}} \sum_{i=1}^{N_{\text{frames}}} d_{\text{perceptual}}(SP\text{ref}_{*i}, SP\text{test}_{*i}) \quad (2)$$

Where $SP\text{ref}_{*i}$ denotes the $i^{\text{th}}$ column of the matrix composed of sound pressure column vectors resulting from preprocessing the reference clip, likewise for $SP\text{test}_{*i}$, and $N_{\text{frames}} = 64$ SP frames. This definition makes the assumption that each frame in a clip contributes equally to its perceived degradation, an assumption that follows from our decision to measure distances between short 1.5-second clips.

The perceptual distance between stimulus points is approximated as the Euclidean distance between transformed (perceptual) points as in Equation 1, following the rationale given in the text.

$$d_{\text{perceptual}}(\vec{x}_{\text{ref}}, \vec{x}_{\text{test}}) = \|f(\vec{x}_{\text{ref}}) - f(\vec{x}_{\text{test}})\| = \sqrt{\sum_{i=1}^{N_{\text{bands}}} \left(f_i(\vec{x}_{\text{ref}}) - f_i(\vec{x}_{\text{test}})\right)^2} \quad (3)$$

where $f_i(\vec{x})$ denotes the $i^{\text{th}}$ component of the stimulus (SP) point $\vec{x}$ mapped to its perceptual representation by function $f$, and $N_{\text{bands}} = 96$ bands. Four methods are proposed for approximating the function mapping stimulus space to perceptual space. The first method is biologically inspired; it models the nonlinear transformations applied to physical stimuli during the mechanical transduction of sound waves into vibration on the basilar membrane. The second method is data-driven; it uses gradient descent to learn a linear mapping from stimulus to perceptual space. The third method uses the experimental data to tune the parameters used to compute the BM transformation matrix. The fourth method approximates processing downstream of the BM transformation as a black-box linear transformation that is optimized using experimental data.

a. BM Model

The BM is essentially a spectrum analyzer, albeit a highly nonlinear and adaptive one. A given part of the BM vibrates the most when stimulated with its characteristic frequency (CF) and less for nearby frequencies, and this motion is detected by the hair cells located on the BM. The inner hair cells transduce vibration into electrical impulses that are passed on to the auditory neurons and undergo further processing by higher levels of the auditory pathway. Outer hair cells detecting vibration increase the sensitivity of nearby regions of the BM, amplifying vibration caused by less intense frequency components [17].

Many attributes of hearing, such as frequency selectivity [18], loudness perception [19], and simultaneous and temporal masking [20] [21], are thought to be established at the level of mechanical transduction in the cochlea. It has been shown that the BM tuning curves are comparable with auditory neural tuning curves in chinchillas [18], suggesting that modeling vibration along the BM might be an appropriate way to represent perceptual vectors in a bottom-up fashion.

We aim to estimate a function of the form $\vec{x}_{BM} = f(\vec{x}_{SP}) = T_{BM}(\vec{x}_{SP})\vec{x}_{SP}$, where the pressure-vibration transformation matrix $T_{BM}(\vec{x}_{SP})$ is stimulus-dependent. While the unavailability of direct measurements of human BM and auditory neural (AN) responses makes it difficult to estimate the human BM response, there are numerous studies demonstrating remarkable similarities between the auditory systems of mammals [22]. Chinchillas are commonly used in hearing research due to ease of access to the inner ear as well as anatomical and physiological similarities to humans [23]. Chinchillas and humans have almost identical hearing and best sensitivity ranges [24], similar auditory filter bandwidths and shapes [25], and similar sharpness of frequency tuning curves [22]. The transformation matrix is constructed based on the results of an experiment by Ruggero et al. measuring the gain (mm/s/Pa) between peak vibration velocity (mm/s) and input sound pressure (Pa) at a single location on the chinchilla BM caused by tones of various loudness levels across a range of frequencies [26]. The gain curves from this paper are replicated empirically (see Supplementary Methods for details), producing the curves shown in Figure 3.[1]

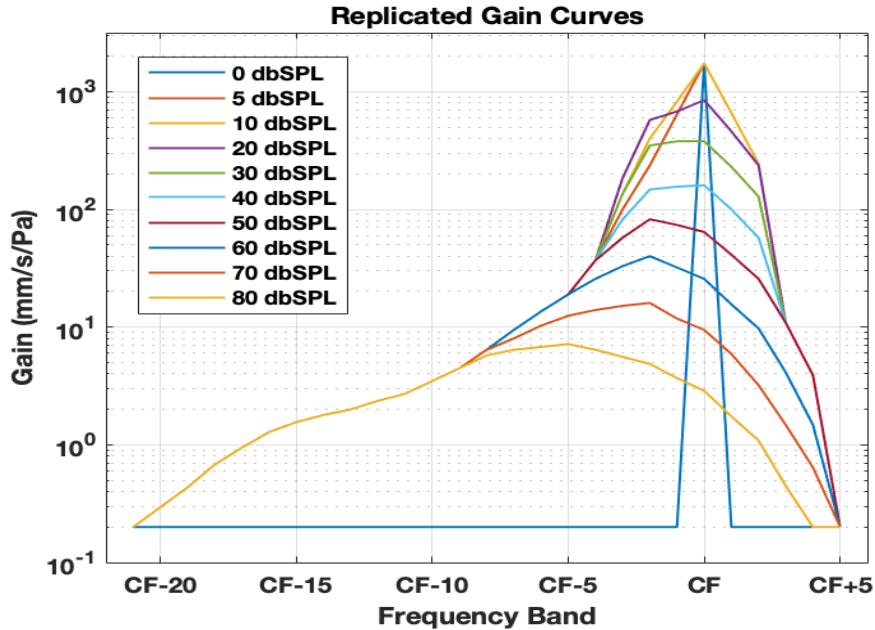

*Figure 3* BM gain curve plot adapted from the Ruggero study using MATLAB

     The curve for each level of input represents the factor by which the equivalent sound pressure is multiplied to produce vibration at a fixed location on the BM when the tone is in a frequency band near the CF of the location in question. For example, consider a stimulus vector with a single 80 dB SPL (0.2 Pa) component present in the 10th frequency band. The BM region whose CF is in the 10th frequency band will vibrate at $(0.2 \text{ Pa})(3 \text{ mm/s/Pa}) = 0.6 \text{ mm/s}$. The same 80 dB component lies 5 bands below the CF of the BM region with CF in the 15th frequency band, so the BM region corresponding to the 15th band will experience a peak vibration velocity of $(0.2 \text{ Pa})(7 \text{ mm/s/Pa}) = 1.4 \text{ mm/s}$. Regions on the BM with CF at or below the 5th band and those with CF at or above the 31st frequency band will vibrate at $(0.2 \text{ Pa})(0.2 \text{ mm/s/Pa}) = 0.04 \text{ mm/s}$; this reflects the component's global effect on vibration

---

[1] Although the gain curve for 0 dB SPL input is not produced in the Ruggero study, for this work the curve used to interpolate gain curves of components below 5 dB SPL is the baseline value everywhere except at CF, where it is the CF gain for 5 dB SPL and 10 dB SPL tones. This ensures the BM transformation matrix is invertible. The other apparent possibility, using a curve that is everywhere the baseline value to interpolate gain curves for components below 5 dB SPL, results in slightly worse performance.

of the entire BM. The differential effect of a component is modeled by subtracting this baseline gain (0.2 mm/s/Pa) from each of the gain curves when constructing the transformation matrix.

For each input component, the appropriate gain curve for the component's SPL is selected by exponentially interpolating between the known gain curves (linearly interpolating between their exponents). Exponential interpolation is chosen because the peaks of the known gain curves are roughly evenly spaced on a logarithmic scale, a reflection of the compressive growth (lower intensity inputs are amplified more than higher intensity inputs; i.e., doubling input less than doubles output) of the BM response. Supposing stimulus vector component $x_i$ has SPL falling somewhere between levels $L_1$ dB SPL and $L_2$ dB SPL ($L_1 \leq \text{SPL}(x_i) \leq L_2$), whose gain curves $10^{\vec{v}_{L1}}$ and $10^{\vec{v}_{L2}}$ are known, the gain curve for $x_i$ would be computed as

$$\text{exponent curve } \vec{v}_{x_i} = \frac{(L_2 - \text{SPL}(x_i))\vec{v}_{L1} + (\text{SPL}(x_i) - L_1)\vec{v}_{L2}}{L_2 - L_1} \quad (4)$$

$$\text{gain curve } \vec{g}_{x_i} = 10^{\vec{v}_{x_i}} - g_{\text{baseline}} \quad (5)$$

where $g_{\text{baseline}} = 0.2$ mm/s/Pa. The gain curve is flipped, then placed in the column of the matrix corresponding to the index of the component, shifting so that the value corresponding to the gain at CF lies on the diagonal of the matrix. This populates the transformation matrix with the appropriate gain values so that each output component is the sum of vibrations produced by the components at its CF band and nearby bands. This is illustrated by a toy example in Figure 4.

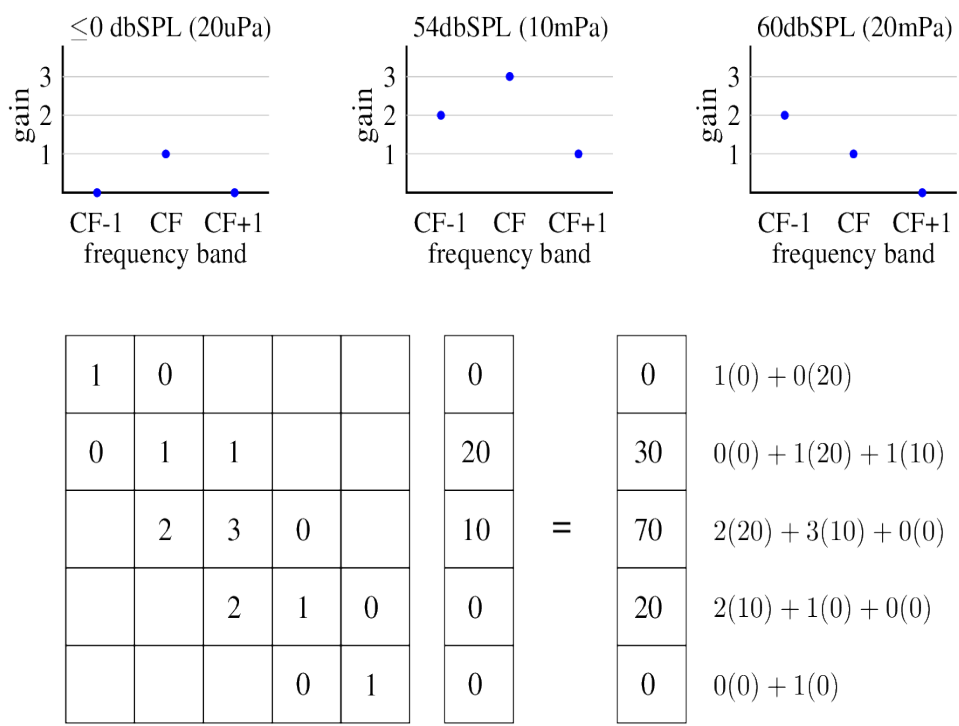

*Figure 4* Toy example demonstrating the construction of the BM gain matrix for a given input stimulus vector. Taking the 10 mPa component as an example, the component is at the CF location of the 3rd BM location, so it experiences a gain of 3 at this location. The same component is 1 band above the CF of the 2nd BM location, so it experiences a gain of 1 at this location, and it is 1 band below the CF of the 4th BM location, so it experiences a gain of 2 at this location

This approximation of BM response might be called "quasilinear" –it imitates many of the nonlinear behaviors of the BM response, including compressive growth (high-intensity inputs are amplified less than low-intensity inputs, increasing the range of audible intensities), frequency selectivity (sharper response for lower intensity inputs), and shifting center frequency (the frequency inducing the largest BM vibration gain for a given intensity level decreases slightly with increasing signal level), but assumes that the BM response to wideband stimuli can be approximated as the superposition of the nonlinear BM response to individual tones. Nonlinear phenomena that emerge as a result of more complex interactions between frequency components – such as distortion products, which are "phantom" tones produced at linear combinations of the frequencies of two simultaneously presented pure tones -- cannot be predicted by a quasilinear description. This approach also ignores the dependence of the precise shape of the gain curves on location along the BM [27]. Despite such limitations, it has been shown that the linear sum of nonlinear responses to tones is a good approximation of the wideband BM response [28]; in the scope of this work, a more detailed approximation of the stimulus-vibration transformation is not attempted. For a given stimulus clip, the transformation matrix is computed for each sound pressure frame (vector) in the clip and used to transform it into a BM vector. The perceptual distance between clips is computed according to Equations 2 and 3 using $f(\vec{x}_{SP}) = T_{BM}(\vec{x}_{SP})\vec{x}_{SP}$.

  b. Data-driven Model

An alternative to inferring a certain form for the perceptual transformation function based on physiology as in the previous method is to assume that most perceptually relevant interactions between frequency components can be captured by a single matrix multiplication and look for the matrix that produces the best predictions of subjective distance ratings. The data-driven approach uses the results of our psychophysical experiment as training data to optimize the weights of a single transformation matrix that is independent of the particular values of the stimulus vector it is applied to. For training, pairs of frames from the reference and degraded clips are used as input. The desired output for a pair of frames is the mean degradation assigned to the clip it is taken from; the simplifying assumption is made that each frame of a short audio clip contributes equally to its perceived degradation. Degradations are rated on a scale of 0 to 100, with 100 corresponding to an imperceptible difference between the reference and degraded clips, so 100 minus the similarity rating is used as the target perceptual distance.

An optimal matrix $M$ is learned by gradient descent, using 1 minus the Pearson correlation between the predicted and subjective distances as the cost function to be minimized. This process is illustrated by Figure 5. $M$ is initialized as the identity matrix, so that training begins with a matrix that leaves input vectors unchanged. Since the values of the transformation matrix are constants, Equation 3 becomes $d_{\text{perceptual}}(\vec{x}_{\text{ref}}, \vec{x}_{\text{test}}) = \|M\vec{x}_{\text{ref}} - M\vec{x}_{\text{test}}\|$.

The cost function is computed using the Pearson correlation
$$\text{corr}(x, y) = r_{xy} = \frac{1}{n-1} \sum_{i=1}^{n} \left(\frac{x_i - \bar{x}}{s_x}\right)\left(\frac{y_i - \bar{y}}{s_y}\right)$$

where $\bar{x}$ is the mean value and $s_x$ the standard deviation of samples $x$, and similarly for $y$.

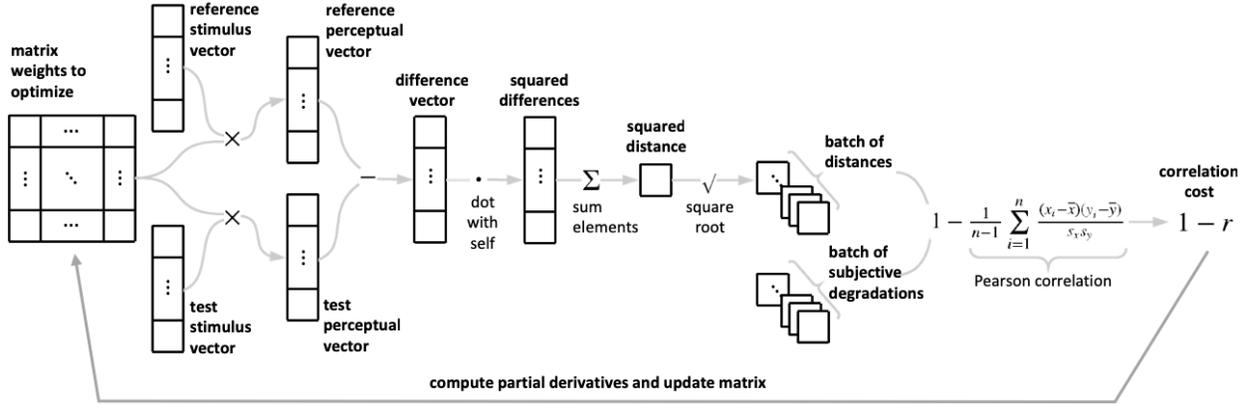

*Figure 5* Diagram of one iteration of the matrix training process. Two vectors being compared are multiplied by the current transformation matrix, and the differences between the resulting vectors are squared, summed, and square rooted, giving a prediction of the perceptual distance between the vectors. The correlation between predicted distance and subjective degradation is computed for a batch of vector pairs. The equations for this process allow for the computation of the rate of change in each output with respect to each input to each step, and the chain rule allows us to chain these expressions to compute the gradient of cost with respect to the matrix weights.

Pearson correlation measures the strength of the linear relationship between variables $x$ and $y$, with $r_{xy} = 1$ indicating a perfect positive linear relationship (i.e., all points $(x_i, y_i)$ lie on a line with positive slope), $r_{xy} = -1$ indicating a perfect negative linear relationship, and $r_{xy} = 0$ indicating no linear relationship. To maximize correlation, the cost function minimized is

$$Cost = 1 - corr(\vec{d}_{\text{predicted}}, \vec{d}_{\text{desired}})$$

where $\vec{d}_{\text{predicted}}$ and $\vec{d}_{\text{desired}}$ are vectors of predicted and actual (subjective) distances for a batch of $B = 512$ frames. The transformation matrix is optimized by batch gradient descent on the cost function. The partial derivatives of the correlation cost function with respect to each element of the matrix are computed from batches of 512 randomly selected training data points, resulting in a vector pointing in the direction of steepest ascent in cost from the current weights, given the current training batch (see Supplementary Methods for an analytical computation of the partial derivatives of cost with respect to matrix weights). The weights are incremented by a small step opposite this direction, given by the negation of the computed gradient multiplied by a step size $\alpha = 0.1$. A new batch is selected, and the process is repeated until the validation cost stops decreasing for some number (say, $N_{\text{converge}} = 1000$) of updates or until a maximum number (say, $N_{\text{maxiter}} = 100000$) of updates have been performed.

For training, the dataset of 480 clips is randomly partitioned into 10 folds of 48 clips each. One fold is set aside as a test set. From the 9 remaining training folds, one is set aside as a validation set for checking whether cost has reached a minimum. Training is repeated, so that each of the 9 training folds is used as the validation set once. Finally, a weighted average of the 9 learned matrices, each weighted by the final correlation between predicted and actual distances for the corresponding validation set, is computed. The test set is used to test the performance of the learned matrix on an unknown set of clips. See Supplementary Discussion for the resulting matrix and discussion.

c. BM Model with Data-driven Parameters

The BM model with data-driven parameters combines the nonlinear modeling capabilities and physiological footing of the BM model with the ability of the data-driven model to leverage experimental data to better predict perceived distances. Instead of directly using the curves from the Ruggero experiment to estimate BM vibration, we use them to initialize the parameters used to compute the gain curve for any given input component value and tune these parameters via gradient descent. As in the fully data-driven method, pairs of frames from reference and test clips and the mean subjective degradations assigned to clip pairs are used for training.

The parameters trained are the base $b$ and the exponent curves $\{\vec{w}_{0\%}, \vec{w}_{5\%}, \ldots, \vec{w}_{100\%}\}$ used to compute the template curves, and a vector $\vec{s}$ of multipliers for each input component to imitate the weighting imposed by the frequency response of the outer and middle ear. They are initialized as follows. First, the SPLs corresponding to the $0^{th}$, $5^{th}$, …, $100^{th}$ percentile values in the stimulus vectors are computed. The initial weights for the exponent curves $\{\vec{w}_{5i\%}\}$ are obtained by interpolating the exponent curves for these SPLs as in Equation 4 and subtracting out the exponent for baseline gain (equivalent to normalizing gain curves by baseline gain). The constant used as the base of the exponential term producing the template curves is initialized as $b = 10$, and the stimulus components' scale factors $\vec{s}$ are initialized as a vector of $N_{\text{bands}} = 96$ ones. The transformation matrix for a given input is constructed as in the BM model, but now with variable parameters controlling the shapes of the curves. The exponent curves for each stimulus component are selected by interpolating between the two exponent curves $\{\vec{w}_{5i\%}, \vec{w}_{5(i+1)\%}\}$ where $i$ is the index of the percentile range the component belongs to. The constant parameter $b$ is raised to the power of each curve. Subtracting 1 results in the gains causing motion relative to baseline velocity[2]. From the matrix-multiplied vectors onward, the correlation cost is computed as in the data-driven model, and the training data is divided as in the data-driven model for cross-validation and model averaging. The training process is illustrated in

---

[2] The gain curves can now be multiplied by the baseline velocity to restore the scale that was changed by subtracting the baseline gain exponent from the exponent curves, but this step is less important because scalar multiplication of the gains is equivalent to the same scalar multiplication of the transformation matrix, which is equivalent to the same scalar multiplication of the perceptual vector and the same scalar multiplication of measured distances (since the 2-norm has the scaling property); i.e., correlation between stimulus distances and perceptual vectors should not be affected.

Figure 6. See Supplementary Methods for a derivation of the partial derivatives of the cost function with respect to each of the parameter weights.

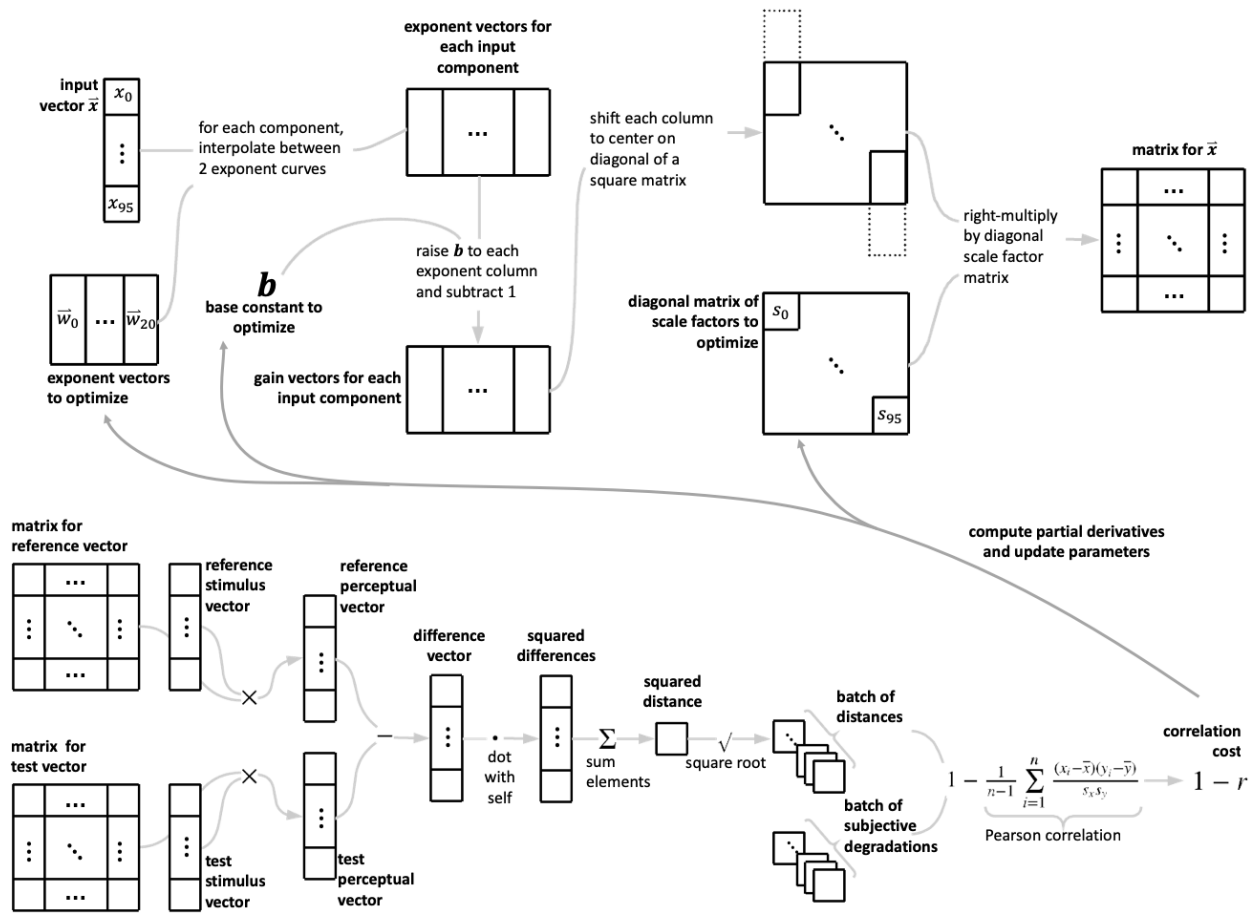

*Figure 6* Diagram of one iteration of the parameter training process. For a given input vector, each component's exponent curve is computed by linear interpolation between curves $\{\vec{w}_{5i\%}\}$. Base $b$ is raised to the power of each exponent curve and placed in the transformation matrix, and each column is scaled by its corresponding component in scale factor vector $\vec{s}$. Besides the fact that there are now two unique transformations matrices for different inputs, the correlation is computed exactly as in Figure 5. Partial derivatives of cost are propagated back beyond the matrix to the parameters used to generate them.

The expectation is that given the approximate form of the mammalian BM response function, a good enough initial setting of parameters to that function, and enough perceptual similarity data, this method should be able to correct some of the inaccuracies introduced by making certain assumptions about the BM response transform. By keeping the exponential form of the function generating gain curves, the compressive nonlinearity of responses is preserved, but the shape and exponential spacing are allowed to vary. Scale factors for each stimulus component are also introduced, allowing for a simple frequency dependency. The resulting function, theoretically, should produce closer approximations to the unknown human BM response curves than the chinchilla BM curves. See Supplementary Discussion for the resulting parameters after training.

d. BM with Data-driven Downstream Model

For this model, we postulate that the functions defining the mappings between the representations of stimuli in consecutive stages of perception are known up to some stage and use optimization to model the subsequent stages, using the output of the known model as input to some optimization process. The mapping that is trained may be parameterized or constrained based on theories or partial knowledge of the downstream processes, or it may be treated as a black box linear transform. We elect to do the latter as a starting point, treating the transformation described by the BM model as the "known" portion of the model and optimizing the weights of a linear map to model downstream processes. This is done by applying the BM transformation to each reference-test pair of stimulus vectors and using the resulting BM vibration vector pairs and corresponding subjective scores as training inputs to the data-driven model, resulting in a matrix $M$. The overall function mapping stimulus points to perceptual points is then $f(\vec{x}_{SP}) = M(T_{BM}(\vec{x}_{SP})\vec{x}_{SP})$. The BM model can be thought of as a model for the peripheral auditory system, and the trained matrix can be thought of as a linear approximation of the central auditory system. See Supplemental Discussion for the resulting downstream transformation matrix.

## III. Results

The most obvious and commonly reported measure of performance in objective audio quality assessment is the correlation between objectively measured quality levels and average quality rating given to a stimulus by subjects of a listening test, or mean opinion scores (MOS). We use the proposed geometry-based framework, exploring several approaches to approximating the perceptual mapping function, for measuring distances in perceptual space to achieve correlations with MOS that are competitive with the audio quality measures PEAQ [4], PEMO-Q [5], and ViSQOLAudio [6].

The approaches to modeling the perceptual map included a physiology-based approach, a data-driven approach, and two hybrid approaches. The physiology-based BM model approximates the matrix that applies the nonlinear transformations occurring during the mechanical transduction of sound pressures to peak velocities of vibration along the BM. The data-driven model approximates bottom-up perception as a black-box linear transformation, optimizing a single matrix to maximize distance-MOS correlation using our experimental dataset. The BM model with data-driven parameters uses the experimental data to tune the parameters used to compute the BM transformation matrix. The BM with data-driven downstream model approximates processing downstream of the BM transformation as a black-box linear transformation that is optimized using experimental data. See Methods for details on each of the approaches as well as on the dataset and procedures used for training and testing. See Supplementary Discussion for scatterplots of measured dissimilarity versus subjective dissimilarity and a comparison of mean and standard deviation of measurements at each degradation level.

Table 1 Comparison of performances of proposed perceptual distance measures and state-of-the-art audio quality measures. Linear correlation is given as Pearson's $r$. Rank correlation coefficients are also given as Spearman's $\rho$ and Kendall's $\tau$ (the functions used by PEMO-Q and ViSQOLAudio to map model outputs to predicted MOS do not affect rank ordering). All p-values are $p \approx 0$ (null hypothesis $H_0: zero\ correlation$).

| Measure | | $r$ | $\rho$ | $\tau$ |
|---|---|---|---|---|
| **BM model** | | 0.7987 | 0.8124 | 0.6197 |
| **Data-driven model** | | 0.8077 | 0.8006 | 0.6042 |
| **BM model with data-driven parameters** | | 0.8298 | 0.7990 | 0.6078 |
| **BM with data-driven downstream model** | | 0.8387 | 0.8093 | 0.6197 |
| **Euclidean measure** | | 0.7032 | 0.7207 | 0.5339 |
| **PEMO-Q** | $1 - PSM_t$ | 0.8131 | 0.8143 | 0.6256 |
| | $5 - MOS$ | 0.5428 | | |
| **ViSQOLAudio** | $1 - NSIM$ | 0.7175 | 0.7702 | 0.5747 |
| | $5 - MOS$ | 0.6041 | | |
| **PEAQ** | Basic | 0.5406 | 0.7365 | 0.5428 |
| | Adv. | 0.4088 | 0.5264 | 0.3549 |

    The correlations with MOS achieved by each approach, as well as those achieved by state-of-the-art measures, are displayed in Table 1. Spearman's and Kendall's rank correlation coefficients are displayed in fairness to PEMO-Q and ViSQOLAudio, which were developed in consideration of the ITU-T recommendation for the evaluation of objective quality prediction models [7], which allows first- and third-order polynomial fittings when measuring the correlation between model output and subjective dissimilarity to account for unknown variations between subjective listening experiments. The table includes the model outputs $PSM_t$ and $NSIM$ of PEMO-Q and ViSQOLAudio, respectively, computed prior to applying the functions trained to map outputs to predicted MOS.

    As expected, the Euclidean measure is not a good predictor of subjective dissimilarity. We submit that nonlinear effects and interactions between stimulus components result in the representations ultimately compared by the perceptual system to differ substantially from the physical stimuli entering the perceptual system, causing the Euclidean distance between stimulus vectors to be a poor metric.

Each of the proposed (non-Euclidean) measures performs competitively with the benchmark measures. All measures except the BM model achieve a Pearson correlation of $r = 0.8$ or better. The fact that the data-naïve BM model works at all suggests that early auditory processing is largely conserved across mammals and that even a rough approximation of the nonlinear BM response captures a significant component of the perceptually relevant transformations of auditory stimuli. The best performance is achieved by the BM model followed by a linear downstream model, supporting the aptness of viewing perception as a series of transformations between representational spaces at different stages of perception.

In our results, the correlations achieved by PEMO-Q and ViSQOLAudio are higher for the unmapped scores; that is, the output of the models before mapping to predicted MOS has better performance. This may indicate that the mapping functions trained for these measures, used out-of-the-box, do not perform well on this dataset. It may similarly be the case that the neural network trained to map PEAQ's model output variables (MOVs) to an objective difference grade (ODG) is not appropriate for this dataset, causing both PEAQ Basic and PEAQ Advanced models to underperform. It should also be noted that PEAQ and PEMO-Q were developed for the measurement of small impairments in audio, so they usually achieve much higher correlations than seen here. The basic version of PEAQ performs better than the advanced version in this case, which is consistent with studies suggesting that the advanced version performs worse for larger degradations [8]. The competing measures were developed to measure impairments in audio degraded by several codecs (e.g. AAC, Opus) in addition to MPEG. It remains to be seen whether the proposed methods can outperform state-of-the-art measures in predicting subjective degradation ratings when a wider variety of codecs is introduced. If our methods remain competitive with these measures when trained on a common dataset on a variety of codecs, they could potentially serve as an alternative to state-of-the-art audio quality measures.

## IV. Conclusions

Based on the conjecture that (bottom-up) perception in general is a smooth transformation of physical stimuli to their perceptual representations, our work introduces a framework for deriving a Riemannian distance metric that more closely tracks perceived dissimilarities between auditory stimuli than the Euclidean metric. Various approaches to modeling the transformation underlying hearing were explored, including techniques informed by physiological processes occurring in the cochlea and techniques informed by psychophysical data.

Each of the proposed measures matches or outperforms against state-of-the-art audio quality measures, encouraging further work addressing the aforementioned caveats to our findings. Our approach is unique from that of existing measures in that our model approximates representations of stimuli that are tangibly present at some level of the perceptual system (e.g. as vibrations along the BM or excitations of auditory neurons) downstream of physical stimulation. The approximated transformations induce a natural definition of distance between stimuli that agrees with actual subjective evaluations of dissimilarity. Our results suggest that perceptual dissimilarity can be interpreted as distance in the mathematic sense.

The disparity in performance between our measures and the Euclidean measure suggests that when treating dissimilarities as distances, the Euclidean distance is a poor selection of metric. We put forward that the dynamic action on and interaction between stimulus components imposed by perception cause the Euclidean metric, which assumes mutual independence between stimulus components, to underperform. We expect that developing more detailed and accurate models of the transformations applied to stimuli during perception will result in still better-performing perceptual distance metrics moving forward.

**Supplementary Methods**

Experiment Interface

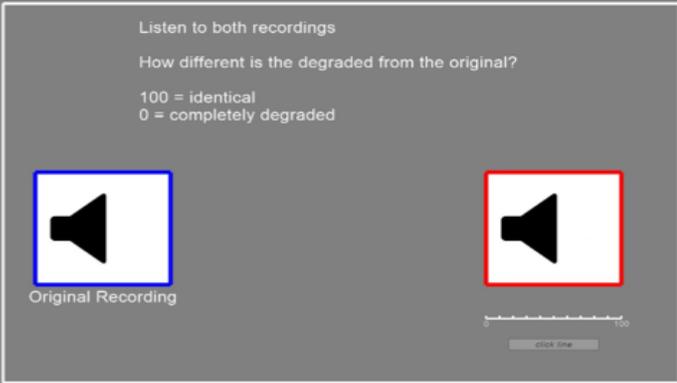

*Supplemental Figure 1 Instructions and rating interface provided to subjects during subjective listening experiment*

Replication of BM Gain Curves

    Graphical methods are used to replicate the curves reported in the Ruggero study [26] in MATLAB. The image containing the plot is opened using `imshow()`, and `ginput()` is used to obtain the pixel locations on the image of the left, right, top, and bottom boundaries of the plot. Each vertical pixel position is converted into the corresponding exponent value on the y axis of the plot.

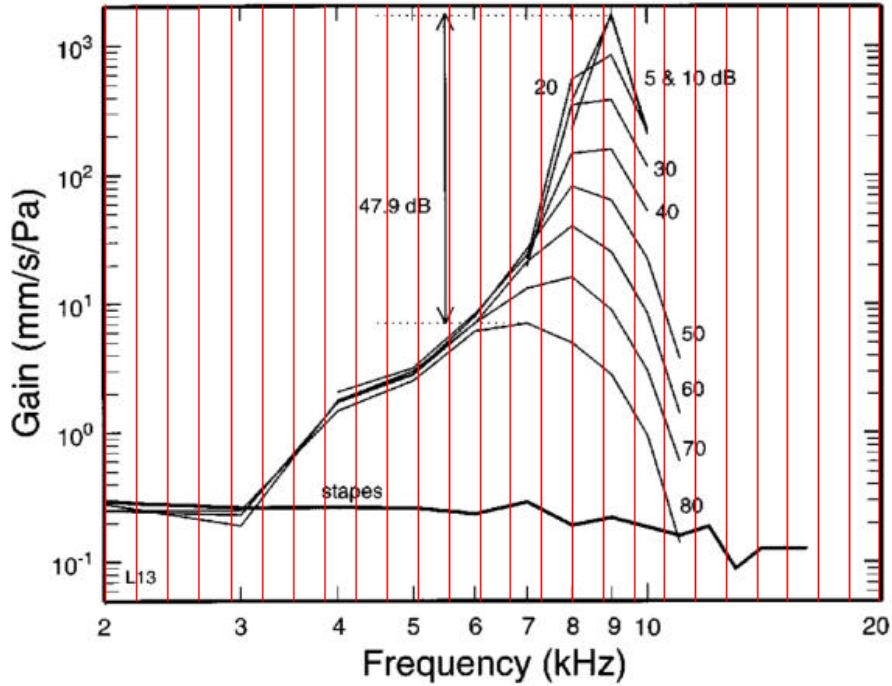

*Supplemental Figure 2* BM gain curve plot from Ruggero study, overlaid with grid lines for point selection

Vertical gridlines corresponding to the frequency band centers in the frequency range of the plot are overlaid on the image. These serve as guide lines for selecting the gain curves at each band center frequency using `ginput()`. The resulting exponent curves are linearly interpolated, giving gain curves varying smoothly with stimulus component level. The curves are used to compute a transformation matrix as described in Methods. For interpolation, it is assumed that the gain at CF remains at the maximum as the gains for neighboring bands fall to zero, ensuring that the transformation matrix is invertible.

Computation of Cost Function Gradient

With Respect to Transformation Matrix
Given a batch of input difference vectors $\vec{x} = SP_{\text{ref}} - SP_{\text{test}}$, the predicted distances $\vec{p}$, and desired distances $\vec{d}$, the gradient of the correlation cost function $C$ with respect to the transformation matrix entries is computed as follows. The correlation is given by

$$corr(\vec{p}, \vec{d}) = \frac{1}{N-1} \sum_{i=1}^{N} \left(\frac{p_i - \bar{p}}{s_p}\right)\left(\frac{d_i - \bar{d}}{s_d}\right)$$

where $\bar{p}$ and $s_p$ are the mean and standard deviation of $\vec{p}$, similarly for $\vec{d}$, and $N$ is the number of samples in the batch; we use $N = B = 512$.

$$s_p = \sqrt{\frac{\sum_{i=1}^{N}(p_i - \bar{p})^2}{N-1}}$$

$$\bar{p} = \frac{\sum_{i=1}^{N} p_i}{N}$$

The correlation cost is then

$$C = 1 - \frac{1}{N-1}\sum_{i=1}^{N}\left(\frac{p_i - \frac{\sum_{i=1}^{N} p_i}{N}}{\sqrt{\frac{\sum_{i=1}^{N}(p_i - \bar{p})^2}{N-1}}}\right)\left(\frac{d_i - \bar{d}}{s_d}\right)$$

For $i = 0, \ldots, N-1$, the partial derivative of the cost w.r.t. the $i^{\text{th}}$ predicted distance $p_i$ is

$$\frac{\partial C}{\partial p_i} = -\frac{1}{N-1}\sum_{i=1}^{N}\left[\left(\delta_{ij} - \frac{1}{N}\right)s_p - \frac{0.5}{s_p}\frac{1}{N-1}2(p_j - \bar{p})\left(\delta_{ij} - \frac{1}{N}\right)\right]\left(\frac{d_i - \bar{d}}{s_d}\right)$$

$$\delta_{ij} = \begin{cases} 1 & i = j \\ 0 & \text{otherwise} \end{cases}$$

For a matrix $Y = [\vec{y}_1, \ldots, \vec{y}_N]$ composed of the perceptual column vectors $\vec{y}_{1,\ldots,N}$ and predicted distances $\vec{p} = [p_1, \ldots, p_N]$ for a batch of $N$ input vectors $\vec{x}_{1,\ldots,N}$, the partial derivative of the cost function w.r.t. the entry in the $i^{\text{th}}$ entry of the $j^{\text{th}}$ column vector is

$$\frac{\partial C}{\partial Y_{ij}} = \frac{\partial C}{\partial p_i}\frac{\partial p_i}{\partial Y_{ij}} \quad \text{where} \quad \frac{\partial p_i}{\partial Y_{ij}} = \frac{1}{2\sqrt{\sum_k Y_{kj}^2}} 2Y_{ij} = \frac{Y_{ij}}{\sqrt{\sum_k Y_{kj}^2}}$$

When $\vec{y}_j$ contains all zeroes (i.e., when $\vec{x}_j$ is all zeroes) this is undefined, but since the desired distance for zero input vectors is 0 they do not contribute to the cost, so we let the partial derivatives equal zero.

Finally, the partial derivatives of the cost function with respect to the weights of the transformation matrix $M$ are given by

$$\frac{\partial C}{\partial M} = \frac{\partial C}{\partial Y} X^T$$

where $X = [\vec{x}_1, \ldots, \vec{x}_N]$ is the matrix composed of the $N$ input vectors.

With Respect to BM Parameters

Compute $\frac{\partial C}{\partial p_i}$, the partial derivatives of cost with respect to the predicted distances, as in the data-driven model. For a matrix $E = Y_{\text{ref}} - Y_{\text{test}}$ containing the differences between a batch of matrix-transformed vectors $Y_{\text{ref}}$ and $Y_{\text{test}}$, and the predicted distances $\vec{p}$ for the batch, the partial derivatives of the cost function with respect to entries of $E$ are given by

$$\frac{\partial C}{\partial E_{ij}} = \frac{\partial C}{\partial p_i}\frac{\partial p_i}{\partial E_{ij}} \quad \text{where} \quad \frac{\partial p_i}{\partial E_{ij}} = \frac{1}{2\sqrt{\sum_k E_{kj}^2}} 2E_{ij} = \frac{E_{ij}}{\sqrt{\sum_k E_{kj}^2}}$$

Since $E = Y_{\text{ref}} - Y_{\text{test}}$, we have

$$\frac{\partial C}{\partial Y_{\text{test}}} = \frac{\partial C}{\partial E}\frac{\partial E}{\partial Y_{\text{test}}} = -\frac{\partial C}{\partial E}$$

We sum the contributions of each sample to the partial derivatives of cost with respect to the scale factor vector $\vec{s}$, the base constant $b$, and the exponent curve weights $\vec{w}_{1,\ldots,22}$ used to construct the gain curves. The contribution of a single pair of input stimuli $\vec{x}_{\text{ref}}, \vec{x}_{\text{test}}$ to the $i^{\text{th}}$ component of $\frac{\partial C}{\partial s}$ is

$$\frac{\partial C}{\partial s_i} = \frac{\partial C}{\partial x_{\text{Sref},i}} x_{\text{ref},i} + \frac{\partial C}{\partial x_{\text{Stest},i}} x_{\text{test},i}$$

where $x_{\text{Sref},i}$ is the $i^{\text{th}}$ component of the scale-factor-vector-weighted input vector $\vec{x}_{\text{Sref}}$ and similarly for $x_{\text{Stest},i}$, and

$$\frac{\partial C}{\partial \vec{x}_{\text{Sref}}} = \frac{\partial C}{\partial \vec{y}_{\text{ref}}} \frac{\partial \vec{y}_{\text{ref}}}{\partial \vec{x}_{\text{Sref}}} \quad \text{where} \quad \frac{\partial \vec{y}_{\text{ref}}}{\partial \vec{x}_{\text{Sref}}}[i,j] = \frac{\partial T_{ij}}{\partial x_{\text{Sref},j}} x_{\text{Sref},j} + T_{ij}$$

where $T_{ij}$ is entry $i,j$ of the transformation matrix computed for that input vector, by the product rule $y_{\text{ref},i} = \sum_{j=1}^{M} T_{ij} x_{\text{Sref},j}$.

$$\frac{\partial T_{ij}}{\partial x_{\text{Sref},j}} = vb^{T_{ij}} (\ln b) \frac{t_{\text{hi},i} - t_{\text{lo},i}}{x_{\text{hi}} - x_{\text{lo}}}$$

since column $j$ of the transformation matrix is $\vec{t}_j = v(b^{\vec{t}} - v)$ where $v$ is the stapes (baseline) velocity and the vector $\vec{t}$ has entries

$$t_k = \frac{(x_{\text{hi}} - x_{\text{Sref},j}) t_{\text{lo},k} + (x_{\text{Sref},j} - x_{\text{lo}}) t_{\text{hi},k}}{x_{\text{hi}} - x_{\text{lo}}} \tag{S1}$$

Here, $x_{\text{hi}}$ and $x_{\text{lo}}$ are the stimulus values of the upper and lower limits of the percentile range $x_{\text{Sref},j}$ falls under. $\vec{t}_{hi}$ and $\vec{t}_{lo}$ are the $j^{\text{th}}$ columns of the two $M \times M$ square matrices formed by moving the two the $\vec{w}_{1,\ldots,22} = \{\vec{w}_{0\%}, \vec{w}_{5\%}, \ldots, \vec{w}_{100\%}\}$ parameter curves corresponding to $x_{\text{hi}}$ and $x_{\text{lo}}$ to the appropriate location along the diagonal. Similarly for $\frac{\partial C}{\partial \vec{x}_{\text{Stest}}}$.

The contribution of a single pair of input stimuli to the partial derivative of cost with respect to the base $b$ of the exponential is given by

$$\sum_{i,j} \left( \frac{\partial C}{\partial T_{\text{ref}}[i,j]} \frac{\partial T_{\text{ref}}[i,j]}{\partial b} + \frac{\partial C}{\partial T_{\text{test}}[i,j]} \frac{\partial T_{\text{test}}[i,j]}{\partial b} \right)$$

where $T_{\text{ref}}$ and $T_{\text{test}}$ are the transformation matrices computed for $\vec{x}_{\text{Sref}}$ and $\vec{x}_{\text{Stest}}$. The $i^{\text{th}}$ column of $\frac{\partial T_{\text{ref}}}{\partial b}$ is given by the element-wise multiplication

$$\frac{\partial T_{\text{ref},*i}}{\partial b} = v\vec{t} \cdot (a^{\vec{t}-1})$$

where $\vec{t}$ is computed for the $i^{\text{th}}$ reference stimulus component using Equation S1. Similarly for $\frac{\partial T_{\text{test}}}{\partial b}$.

The contribution of a single pair of input stimuli to the partial derivative of cost with respect to the weights $\{\vec{w}_{0\%}, \vec{w}_{5\%}, \ldots, \vec{w}_{100\%}\}$ are computed starting with the contribution to the derivative with respect to the entries of the multiplying matrices $T_{\text{ref}}$ and $T_{\text{test}}$ as

$$\frac{\partial C}{\partial T_{\text{ref}}} = \frac{\partial C}{\partial \vec{x}_{\text{Sref}}} (\vec{x}_{\text{Sref}})^T$$

and similarly for $\frac{\partial C}{\partial T_{\text{test}}}$. We iterate through the percentile ranges, keeping only the entries of $\frac{\partial C}{\partial T_{\text{ref}}}$ that contribute to the weights for the current percentile range and zeroing the rest; i.e., we zero $i^{\text{th}}$ columns for which $x_{Sref,i}$ falls outside the current percentile range. Then the contribution to the partial derivative becomes

$$\frac{\partial C}{\partial w_{lo,k}} = \sum_{i,j} \frac{\partial C}{\partial T_{ref}[i,j]} \frac{\partial T_{ref}[i,j]}{\partial w_{lo,k}} \quad \text{where} \quad \frac{\partial T_{ref}[i,j]}{\partial w_{lo,k}} = \ln b \, \frac{x_{hi} - x_{Sref,j}}{x_{hi} - x_{lo}} v\big(b^{\vec{t}[k]}\big)$$

where $\vec{t}[k]$ is the element corresponding to $w_{lo,k}$ in the vector $\vec{t}$, which is generated using Equation S1 by shifting a weighted sum of $\vec{w}_{lo}$ and $\vec{w}_{hi}$ to the proper location. Similarly for $\vec{w}_{hi}$, but $x_{hi} - x_{Sref,j}$ is replaced by $x_{Sref,j} - x_{lo}$. $\vec{x}_{Stest}$ contributes similarly.

For the next percentile range up, $\vec{w}_{hi}$ becomes the new $\vec{w}_{lo}$, and the curve corresponding to the upper boundary of the new percentile range becomes $\vec{w}_{hi}$. The sum of the contributions to $\frac{\partial C}{\partial \vec{w}_{1,...,22}}$ by each $\vec{x}_{ref}$ and $\vec{x}_{test}$ in the batch is the total gradient with respect to the weight curves $\vec{w}_{1,...,22}$.

**Supplementary Discussion**

Statistics for Listener Evaluations of Dissimilarity by Degradation Level

Supplemental Table 1 and Supplemental Figure 3 show statistics for the ratings given to clips in the experiment described in Methods. Each degradation of each of the 480 clips in our dataset was evaluated by 3 subjects, resulting in a single mean opinion score (MOS) for each degradation. The mean and standard deviation of MOS for the 480 clips are displayed for each degradation level.

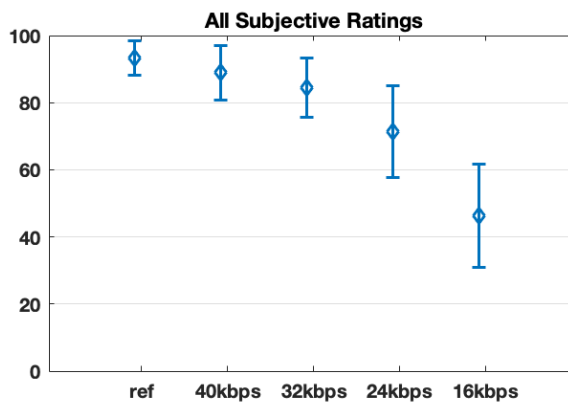

| Degradation Level | mean MOS | std. dev. MOS |
|---|---|---|
| Reference | 93.33 | 5.11 |
| 40 kbps | 88.90 | 8.19 |
| 32 kbps | 84.48 | 8.80 |
| 24 kbps | 71.34 | 13.58 |
| 16 kbps | 46.42 | 15.38 |

*Supplemental Table 1* Average and standard deviations of the MOS assigned to clips by users for each degradation level

*Supplemental Figure 3* Error bar plots for statistics shown in Supplemental Table 1

Scatterplots for Proposed and State-of-the-Art Measures

The scatterplots displayed in Figure 4 show the model-predicted dissimilarity against $100 - MOS$. Points plotted in gray are measurements for which the test clip was the undegraded reference clip. Pearson's $r$, Spearman's $\rho$ and Kendall's $\tau$ are displayed for all reference-test pairs for each of the 50 clips in the test set. The accompanying errorbar plots in Figure 4 display the mean and standard deviation error bars of scores for all pairs in the test set for each MPEG bitrate.

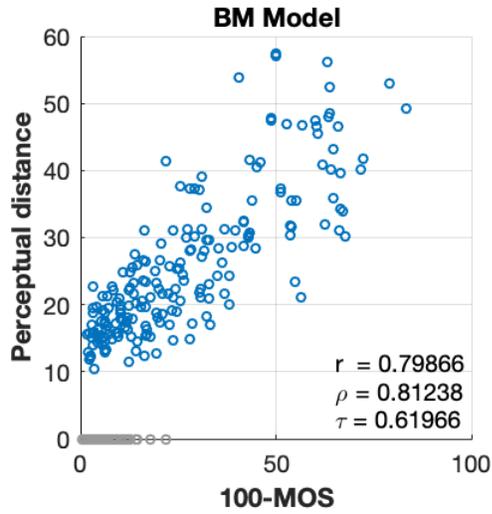

*Supplemental Figure 4 a*

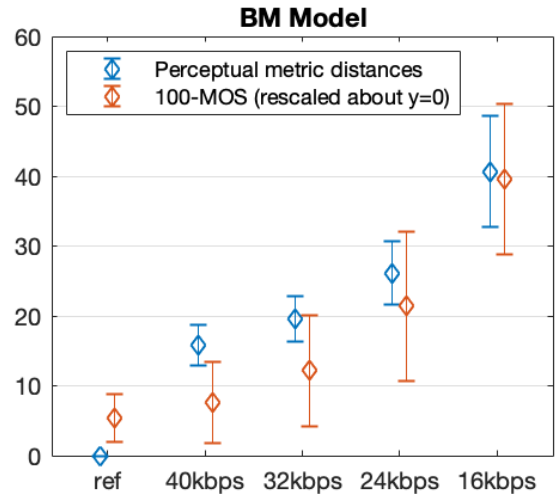

*Supplemental Figure 4 b*

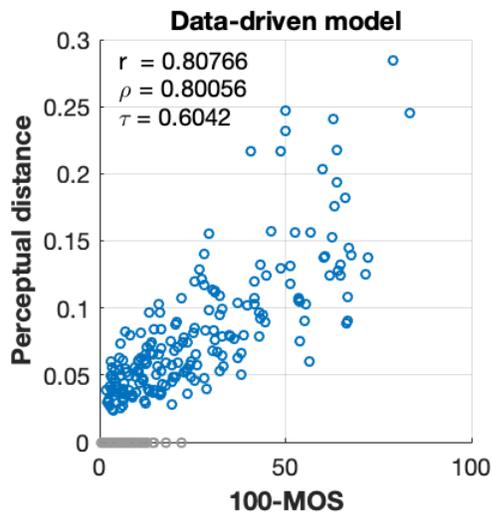

*Supplemental Figure 4 c*

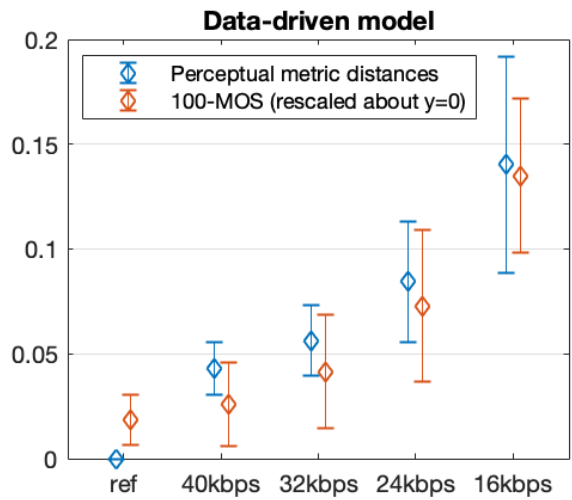

*Supplemental Figure 4 d*

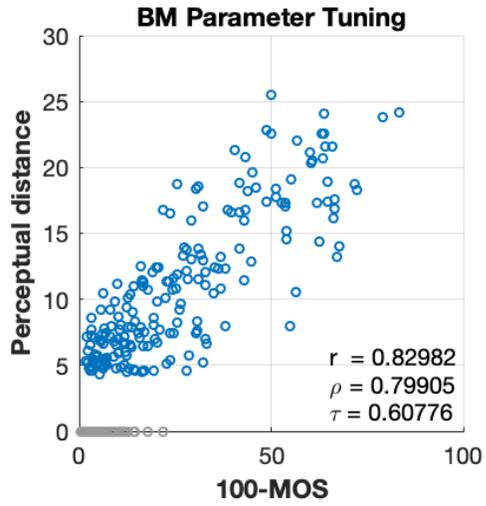

*Supplemental Figure 4 e*

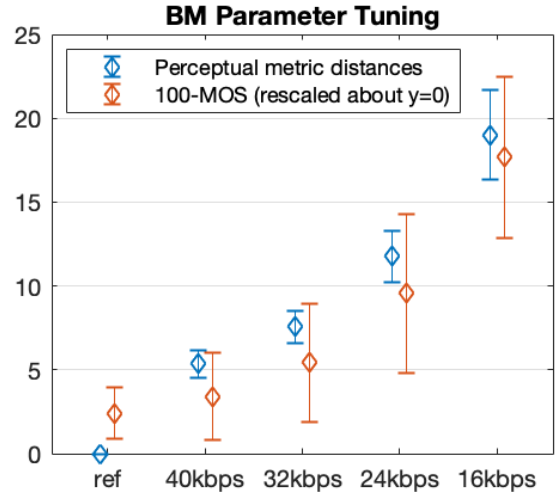

*Supplemental Figure 4 f*

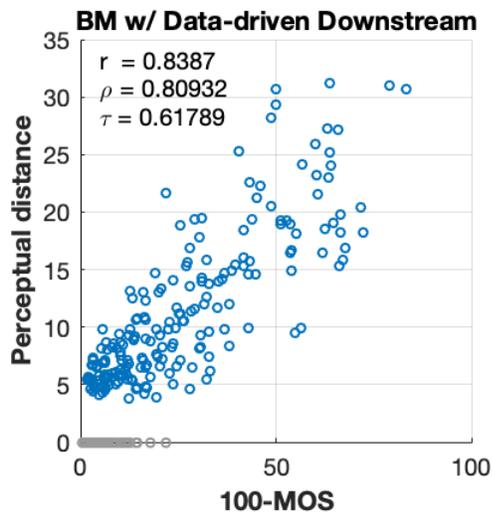

*Supplemental Figure 4 g*

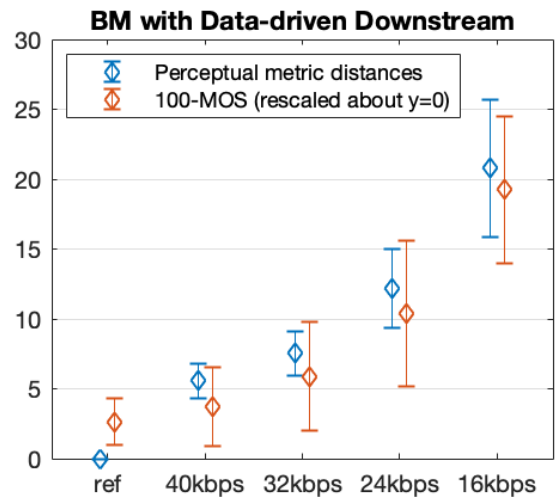

*Supplemental Figure 4 h*

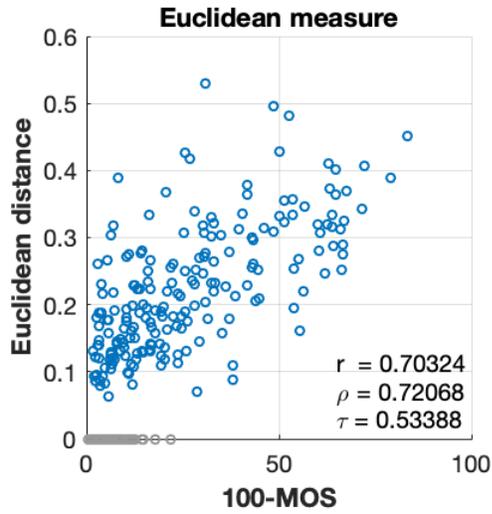

*Supplemental Figure 4 i*

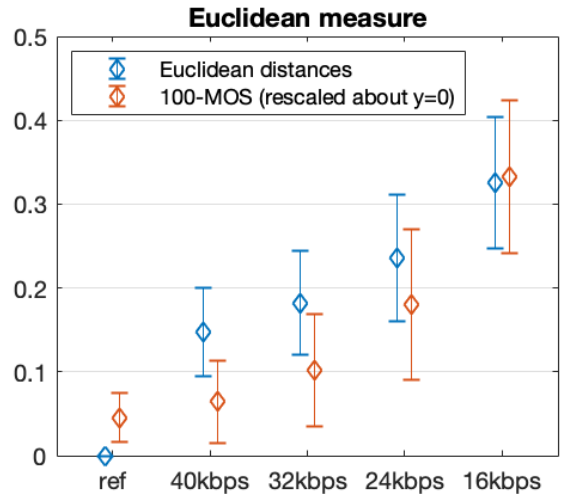

*Supplemental Figure 4 j*

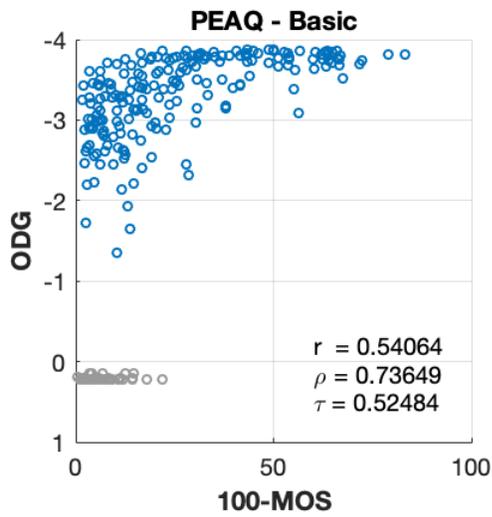

*Supplemental Figure 4 k*

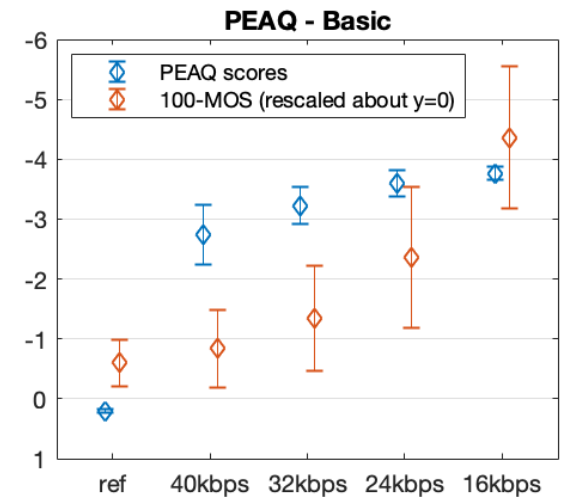

*Supplemental Figure 4 l*

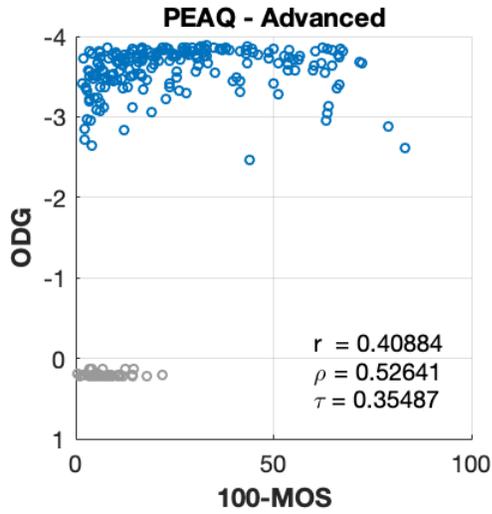

*Supplemental Figure 4 m*

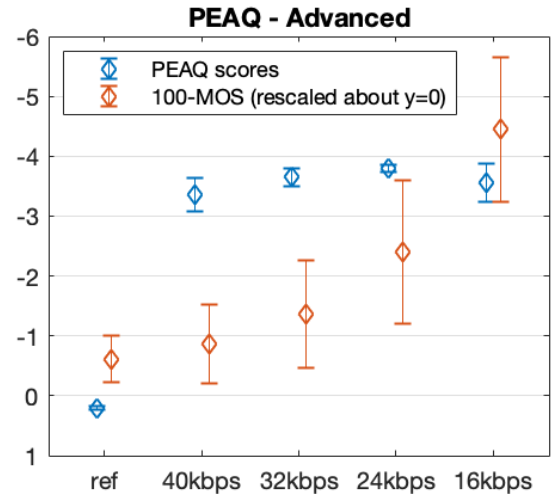

*Supplemental Figure 4 n*

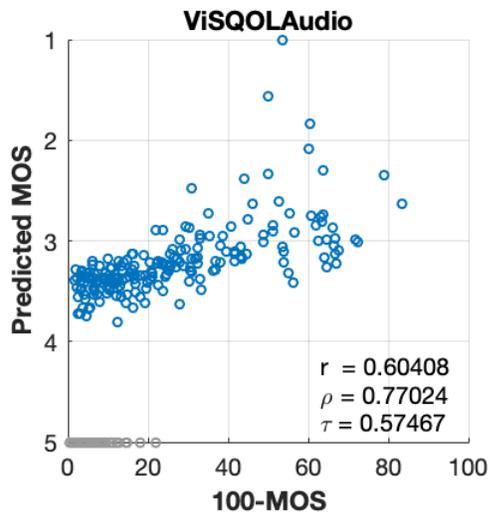

*Supplemental Figure 4 o*

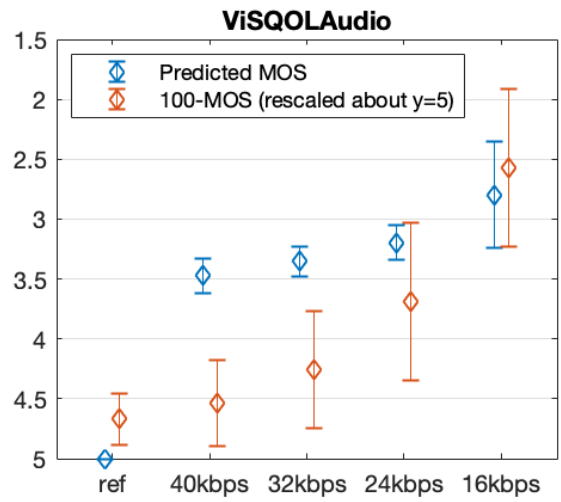

*Supplemental Figure 4 p*

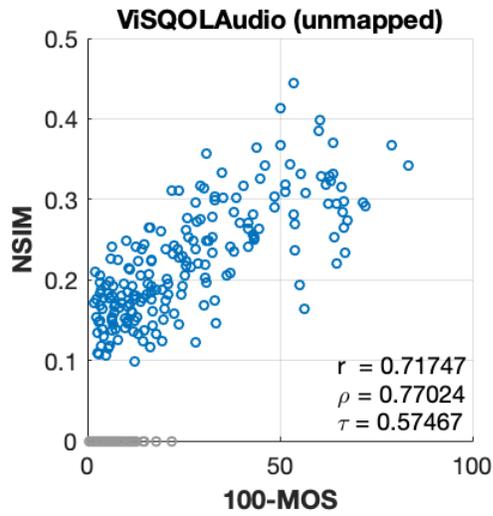

*Supplemental Figure 4 q*

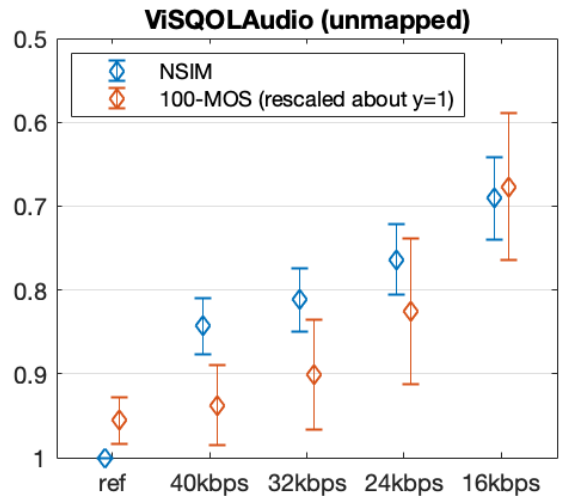

*Supplemental Figure 4 r*

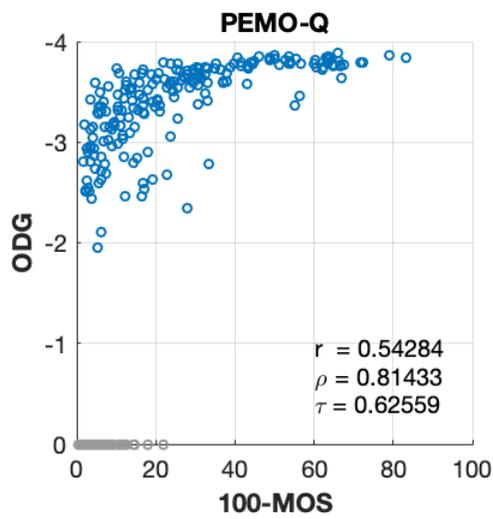

*Supplemental Figure 4 s*

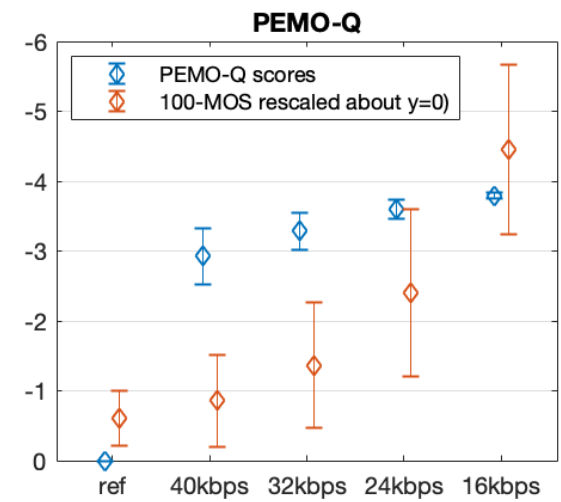

*Supplemental Figure 4 t*

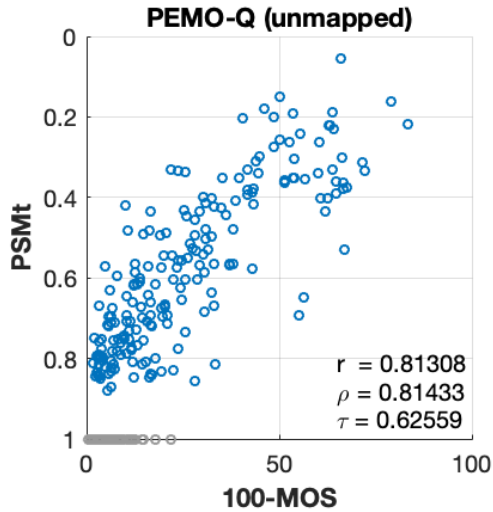 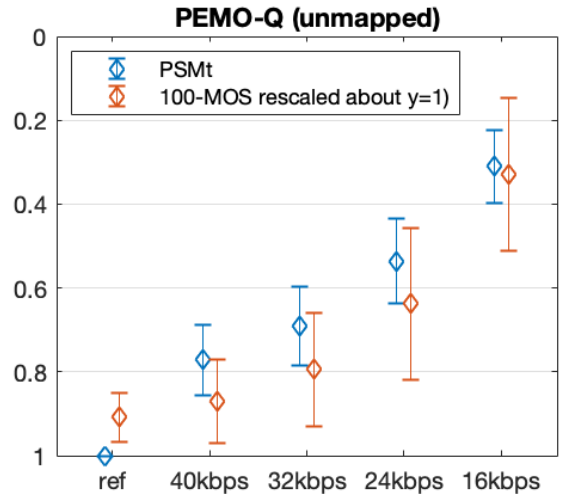

*Supplemental Figure 4 u*          *Supplemental Figure 4 v*

*Supplemental Figure 4* Left column displays measure vs MOS scatterplots; right column displays error bars for measured and subjective dissimilarity in each degradation level.

Results of Training for Data-driven Model

The heatmap in Supplemental Figure 5a shows the final matrix obtained using the data-driven model. Inspecting the heatmap of the matrix, we can still make out the diagonal, but it is "sharpened," with the positive values along the diagonal surrounded by slightly negative values, at least on the left side of the matrix. The right side of the matrix exhibits striation patterns which are not oriented about the diagonal.

A few columns are sampled from the matrix and shown in Supplemental Figure 5b. These columns can be thought of as flipped and shifted gain curves for the corresponding input frequency components; column $i$ shows how frequency component $i$ contributes to each of the output components. Interestingly, the columns selected from the left side of the Jacobian appear similar to differences of Gaussians (DoGs), the two-dimensional version of which is often used to model the receptive fields when lateral inhibition is present, as in the case of retinal ganglion cells in computer vision [29]. The effect of such a column is to contribute some quantity proportional to the input component to the corresponding output component, and subtract some smaller proportion of the input component from nearby output components. Taken together, these columns have the effect of sharpening peaks in the input stimulus vector.

Columns taken from the right side of the matrix take on a more irregular shape, with the peak along the diagonal still present, but accompanied by bumps and dips that vary with frequency band rather than relative to the diagonal. This may be partly explained by the fact that zero-value components do not contribute to the gradient, and these high-frequency components are zero more often than not. The average component value in each frequency band is shown in Supplemental Figure 5c. We may suspect that a high occurrence of zero inputs and a general bias toward small component magnitudes in higher frequency bands causes the weights of the right side of the matrix, which multiply the higher frequency components going into the output vectors, not to take on a meaningful form during optimization.

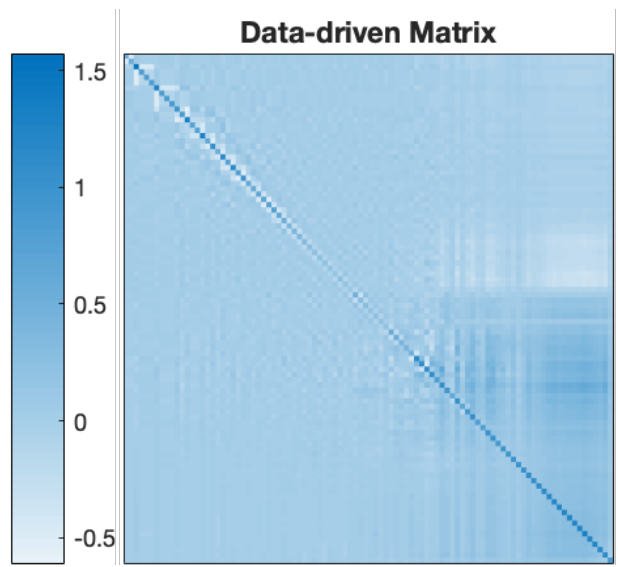

*Supplemental Figure 5 a*

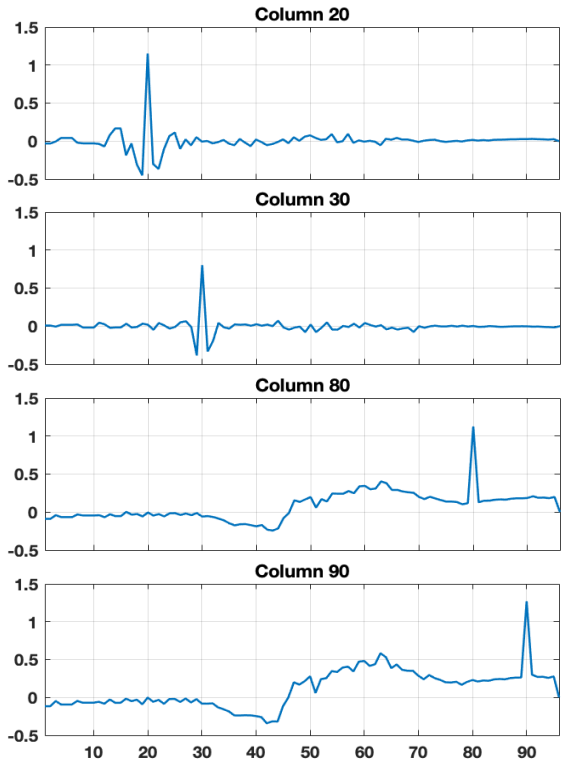

*Supplemental Figure 5b*

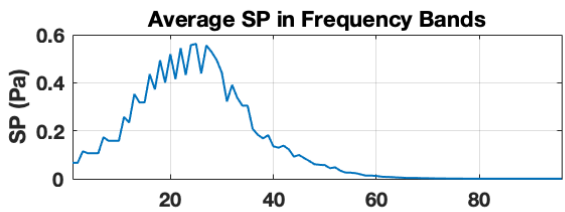

*Supplemental Figure 5 c*

*Supplemental Figure 5* Results of training data-driven model: heatmap (a), selected columns (b), and average value of each input component (c)

Results of Training for BM Model with Data-driven Parameters

       The plots in Supplemental Figure 6 display the parameters used to compute the BM transformation matrix, before and after training. The base of the exponent increases from its initial value of 10 to 10.5. The exponent curves that are interpolated, used to raise the base to the appropriate power, and flipped to compute gain curves for a given input component are plotted. Training causes the curves to become sharper and increases the leftward shift of the characteristic frequency at high intensities relative to the characteristic frequency at low intensities.

       It was hypothesized that the scale factors would imitate outer and middle ear's amplification/attenuation of certain frequencies. Comparing the trained scale factors with PEAQ's outer and middle ear transfer function, plotted along with the scale factors against frequency, suggests this may have occurred to an extent.

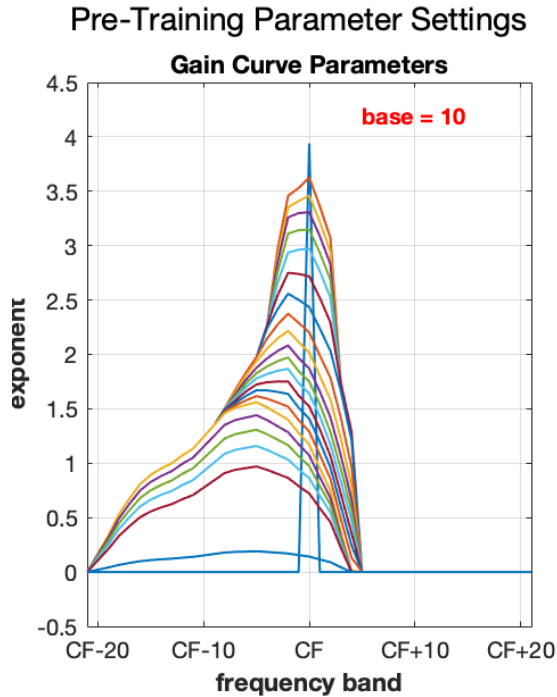
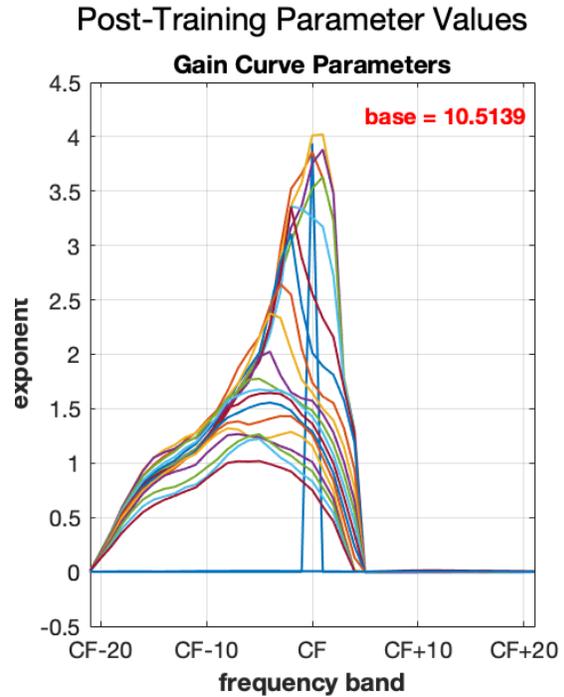
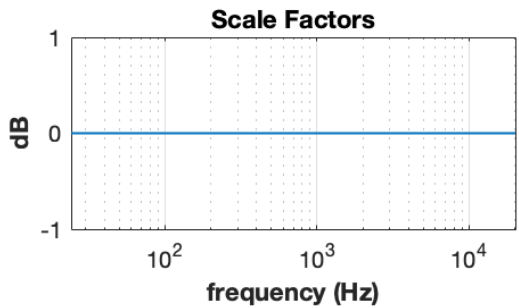
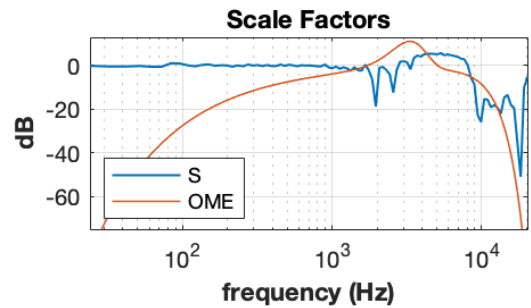

*Supplemental Figure 6 a* Parameter values before training.

*Supplemental Figure 6 b* Parameter values after training. OME shows PEAQ's outer and middle ear transfer function for comparison with learned scale factors.

*Supplemental Figure 6* Parameter values before and after training. The base constant is to be raised to the power of an interpolated exponent curve to compute a gain curve for a component. The scale factors are plotted against the center frequencies of the frequency bands they are responsible for scaling.

Results of Training for BM with Data-driven Downstream Model

The heatmap in Supplemental Figure 7a displays the matrix learned by the downstream modeling method. This is the matrix optimized using the same process as the data-driven method, but using the output of the BM model as input. Like the matrix learned by the purely data-driven method, it has a "sharpening" effect, but its rows and columns have a more regular form, and the striation patterns appearing in the matrix for the purely data-driven method are not observed. Columns 20, 30, 80, and 90 are plotted in Supplemental Figure 7b for comparison with the same columns shown for the data-driven method.

As is often the case for results in machine learning, it is difficult to say why this happens, but one possibility is that the pre-amplification of the mostly low-intensity components in higher frequency bands by the BM model result in the weights for these components (i.e., the right side of the downstream matrix) having a larger influence on the final output vector than they would have without pre-amplification. Without pre-amplification, weights on the right side of the matrix contribute less to each output component because the components they multiply tend to be so small, causing training to be more noisy.

If interpreted as a black-box model for transformations occurring downstream of the stimulus-to-BM transform, this matrix may be some indication of the overall effect of the neural circuits in the cochlear nucleus , which incorporate both excitatory and lateral inhibitory processes.

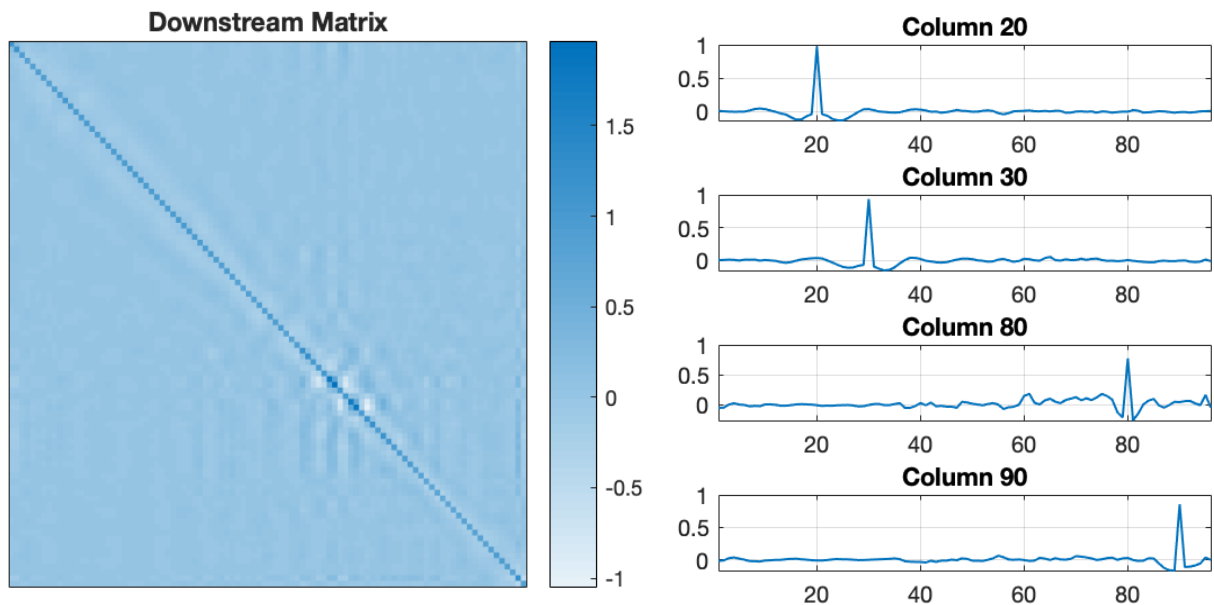

*Supplemental Figure 7 a*

*Supplemental Figure 7 b*

*Supplemental Figure 7* Results of training BM with data-driven downstream model. Heatmap showing trained portion of model (a), selected columns (b)

**Acknowledgements**
This work was supported in part by grants N00014-15-1-2132 and N00014-16-1-2359 from the Office of Naval Research.

**Author Contributions**
A.R. and R.G. developed the original concept of a non-Euclidean geometric framework for perceptual similarity. D.S. contributed to the theoretical framework. E.B. led initial efforts to apply the framework for image similarity, aided the adaptation of the framework to audio, collected auditory stimuli, and created the dataset used for this work. E.C. and A.B. designed and conducted the experiments. S.O. developed the perceptual distance measures for auditory stimuli and wrote the manuscript with assistance from E.B., D.S., L.R., and R.G.